\documentclass[%
aps,
prb,
 amsmath,amssymb,
 superscriptaddress,
preprint,%
nobibnotes,
longbibliography
]{revtex4-1}

\usepackage{graphicx}
\usepackage{dcolumn}
\usepackage{bm}
\usepackage{ulem}
\usepackage[utf8]{inputenc}
\usepackage[T1]{fontenc}
\usepackage{mathptmx}
\usepackage{textcomp} 
\usepackage{nicefrac}
\usepackage{xcolor}
\usepackage{gensymb}
\usepackage{multibib}
\usepackage[margin=3cm]{geometry}
\usepackage[utf8]{inputenc}
\usepackage{physics}
\usepackage{amsmath}

\DeclareMathAlphabet{\pazocal}{OMS}{zplm}{m}{n}
\newcommand{\Tb}{\pazocal{T}}

\newcommand{\Da}{\mathcal{D}}
\newcommand{\Ea}{\mathcal{E}}

\usepackage{lineno}

\def\bibsection{\section{\refname}} 

\usepackage{xr}
\makeatletter
\newcommand*{\addFileDependency}[1]{
  \typeout{(#1)}
  \@addtofilelist{#1}
  \IfFileExists{#1}{}{\typeout{No file #1.}}
}
\makeatother

\newcommand*{\myexternaldocument}[1]{%
    \externaldocument{#1}%
    \addFileDependency{#1.tex}%
    \addFileDependency{#1.aux}%
}
\myexternaldocument{methods}

\begin{document}



\title[Probing correlations in the exciton landscape of a moiré heterostructure]{Probing correlations in the exciton landscape of a moiré heterostructure}


\author{Jan Philipp Bange} %
\address{I. Physikalisches Institut, Georg-August-Universit\"at G\"ottingen, Friedrich-Hund-Platz 1, 37077 G\"ottingen, Germany}

\author{David Schmitt} %
\address{I. Physikalisches Institut, Georg-August-Universit\"at G\"ottingen, Friedrich-Hund-Platz 1, 37077 G\"ottingen, Germany}

\author{Wiebke Bennecke} %
\address{I. Physikalisches Institut, Georg-August-Universit\"at G\"ottingen, Friedrich-Hund-Platz 1, 37077 G\"ottingen, Germany}

\author{Giuseppe Meneghini} 
\address{Fachbereich Physik, Philipps-Universit{\"a}t, 35032 Marburg, Germany}

\author{AbdulAziz AlMutairi} 
\address{Department of Engineering, University of Cambridge, Cambridge CB3 0FA, U.K.}

\author{Kenji Watanabe} %
\address{Research Center for Functional Materials, National Institute for Materials Science, 1-1 Namiki, Tsukuba 305-0044, Japan}

\author{Takashi Taniguchi} %
\address{International Center for Materials Nanoarchitectonics, National Institute for Materials Science, 1-1 Namiki, Tsukuba 305-0044, Japan}

\author{Daniel Steil} %
\address{I. Physikalisches Institut, Georg-August-Universit\"at G\"ottingen, Friedrich-Hund-Platz 1, 37077 G\"ottingen, Germany}

\author{Sabine Steil} 
\address{I. Physikalisches Institut, Georg-August-Universit\"at G\"ottingen, Friedrich-Hund-Platz 1, 37077 G\"ottingen, Germany}

\author{R. Thomas Weitz} %
\address{I. Physikalisches Institut, Georg-August-Universit\"at G\"ottingen, Friedrich-Hund-Platz 1, 37077 G\"ottingen, Germany}
\address{International Center for Advanced Studies of Energy Conversion (ICASEC), University of Göttingen, Göttingen, Germany}

\author{G.~S.~Matthijs~Jansen} %
\address{I. Physikalisches Institut, Georg-August-Universit\"at G\"ottingen, Friedrich-Hund-Platz 1, 37077 G\"ottingen, Germany}

\author{Stephan Hofmann} 
\address{Department of Engineering, University of Cambridge, Cambridge CB3 0FA, U.K.}

\author{Samuel Brem} 
\address{Fachbereich Physik, Philipps-Universit{\"a}t, 35032 Marburg, Germany}

\author{Ermin Malic} 
\address{Fachbereich Physik, Philipps-Universit{\"a}t, 35032 Marburg, Germany}
\address{Department of Physics, Chalmers University of Technology, Gothenburg, Sweden}

\author{Marcel Reutzel} \email{marcel.reutzel@phys.uni-goettingen.de}%
\address{I. Physikalisches Institut, Georg-August-Universit\"at G\"ottingen, Friedrich-Hund-Platz 1, 37077 G\"ottingen, Germany}

\author{Stefan Mathias} \email{smathias@uni-goettingen.de}%
\address{I. Physikalisches Institut, Georg-August-Universit\"at G\"ottingen, Friedrich-Hund-Platz 1, 37077 G\"ottingen, Germany}
\address{International Center for Advanced Studies of Energy Conversion (ICASEC), University of Göttingen, Göttingen, Germany}

\begin{abstract}

Excitons are two-particle correlated bound states that are formed due to Coulomb interaction between single-particle holes and electrons. In the solid-state, cooperative interactions with surrounding quasiparticles can strongly tailor the exciton properties and potentially even create new correlated states of matter. It is thus highly desirable to access such cooperative and correlated exciton behavior on a fundamental level. Here, we find that the ultrafast transfer of an exciton's hole across a type-II band-aligned moiré heterostructure leads to a surprising sub-200-fs upshift of the single-particle energy of the electron being photoemitted from the two-particle exciton state. While energy relaxation usually leads to an energetic downshift of the spectroscopic signature, we show that this unusual upshift is a clear fingerprint of the correlated interactions of the electron and hole parts of the exciton quasiparticle. In this way, time-resolved photoelectron spectroscopy is straightforwardly established as a powerful method to access exciton correlations and cooperative behavior in two-dimensional quantum materials. Our work highlights this new capability and motivates the future study of optically inaccessible correlated excitonic and electronic states in moiré heterostructures.

\end{abstract}

\maketitle

Correlated interactions of quasiparticles in solids lie at the heart of condensed-matter physics. They are fundamental to symmetry broken phases such as Mott and Wigner crystals~\cite{Regan20nat,Smolenski21nat,Brem22Wigner}, and promise emergent material properties for the next generation technological devices~\cite{Liang20advmat}. A prime example of a strongly correlated quasiparticle is the Coulomb-bound electron-hole pair, i.e., an exciton. As in the case of a hydrogen atom, the exciton's properties are described by its quantum number, its binding energy and its Bohr radius~\cite{Wang18rmp}. For low-dimensional materials, these key parameters can be significantly altered by cooperative interactions with surrounding quasiparticles~\cite{Chernikov14prl,He14prl}. In order to study such cooperative and emergent correlated behavior, artificial stacks of two-dimensional transition metal dichalcogenides (TMDs) have been shown to provide an exceptional playground for manipulating exciton properties. Examples include the creation of interlayer excitons~\cite{Lee14natnano,Hong14natnano,Ceballos14acsnano,Merkl19natmat}, the confinement of excitons in a moiré potential well~\cite{Seyler19nat, Alexeev19nat, Tran19nat}, the creation of correlated interlayer exciton insulators~\cite{Ma21nat,Zhang22natphys} and exciton crystals~\cite{Slobodkin20prl,Sigl20prr}, and even the stabilization of Bose-Einstein condensates~\cite{Wang19nat}. 

\begin{figure}[hbt!]
    \centering
    \includegraphics[width=.5\linewidth]{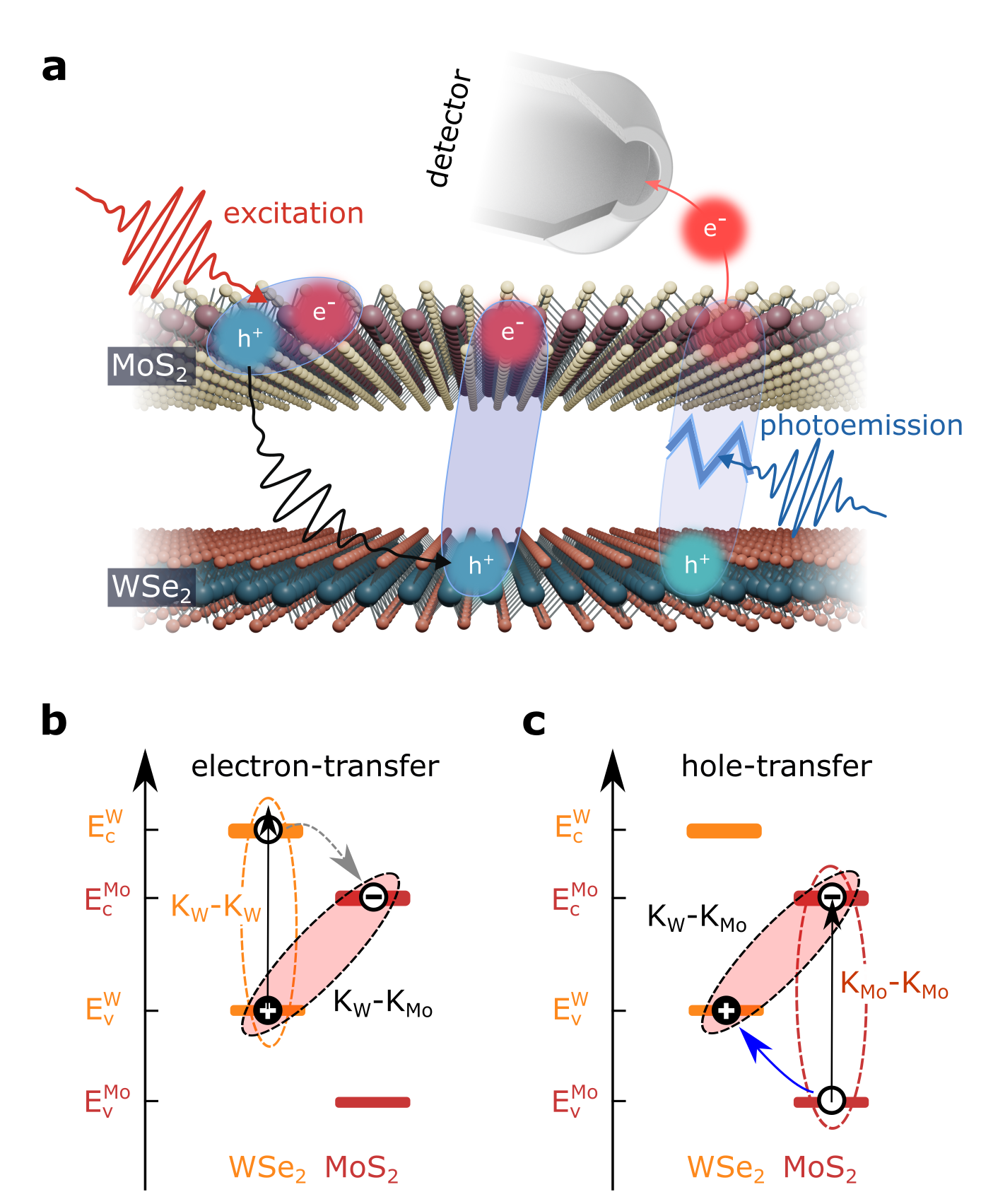}
    \caption{\textbf{Probing Coulomb correlated electron-hole pairs and their femtosecond dynamics using momentum microscopy.}
    (\textbf{a}) Schematic illustration of the photoemission process from excitons. Visible femtosecond light pulses (red) are used to optically excite bright excitons that fully reside in the MoS$_2$ monolayer. The transfer of the hole-component into the WSe$_2$ monolayer leads to the formation of charge separated interlayer excitons (black arrow). A time-delayed extreme ultraviolet laser pulse (blue) breaks the exciton species; single-particle electrons are detected in the photoelectron analyzer and single-particle holes remain in the WSe$_2$ monolayer.
    (\textbf{b,c}) Single-particle energy-level alignment of the valence and conduction bands (v, c) of MoS$_2$ and WSe$_2$. K$_{\rm W}$-K$_{\rm Mo}$ excitons are formed due to interlayer charge transfer of the exciton's hole or electron, respectively, from intralayer K$_{\rm Mo}$-K$_{\rm Mo}$ or K$_{\rm W}$-K$_{\rm W}$ excitons. Note that in (\textbf{c}), the electron-contribution to the exciton remains rigid in the conduction band minimum of MoS$_2$ during the hole transfer process. In the abbreviation of the excitons, the capital letters and the subscripts denote the valley (K, $\Sigma$, $\Gamma$) and the layer (W, Mo) where the exciton's hole (first letter) and electron (second letter) are localized. }
\end{figure}

It is of fundamental importance to obtain insight into the energy landscape and the dynamics of the correlated electron-hole pairs~\cite{Jin18natnano}. In TMD semiconductors, momentum-indirect and spin-forbidden excitons play a significant role, but are mostly~\cite{Poellmann15natmat,Merkl19natmat} inaccessible using all-optical experimental techniques~\cite{Malic18prm,mueller18,Perea22apl}. Recent time- and angle-resolved photoelectron spectroscopy (trARPES) experiments have been shown to be a powerful technique to fill this gap and to simultaneously probe the energy landscape and dynamics of optically bright and dark excitons in monolayer~\cite{Madeo20sci, Wallauer21nanolett, kunin23prl} and twisted bilayer~\cite{Schmitt22nat, Karni22nat} TMDs (Fig.~1a). Importantly, when using photoelectron spectroscopy, there is a fundamental aspect that needs to be considered (Fig.~1a): in the photoemission process, the Coulomb correlation between the electron- and the hole-component of the exciton is broken. This is because a single-particle photoelectron is collected with the detector and a single-particle hole remains in the material~\cite{Weinelt04prl,Perfetto16prb,Rustagi18prb,Dong20naturalsciences,Man21sciadv}. As a consequence, trARPES provides seemingly natural access to the electron contribution of the correlated quasiparticle and thus to the charge-transfer of the exciton's electron across the type-II band aligned interface~\cite{Schmitt22nat} (Fig.~1b). However, to this day, it has not been shown that trARPES can be applied to the inverse process, i.e., the charge-transfer dynamics of the exciton's hole across the TMD interface (Fig.~1c). Hence, exactly this hole-transfer channel in the interlayer exciton formation process remains so far elusive~\cite{Lee14natnano,Hong14natnano,Ceballos14acsnano,Zhu17nanolett, Zimmermann21acsnano, Policht21nanolett,Wang21nanolett,Jin18natnano}.

Here, we demonstrate how the Coulomb interaction between the electron- and the hole-components of the intra- and interlayer excitons facilitates the study of the ultrafast hole-transfer mechanism in a twisted WSe$_2$/MoS$_2$ heterostructure. Interestingly, we experimentally observe an increase of the exciton's photoelectron energy upon the hole-transfer process across the interface. This is surprising at first, because the electron remains rigid in the conduction band minimum during this hole-transfer process (Fig.~1c), and also because any relaxation mechanism is typically expected to cause an overall decrease of the measured electronic quasiparticle energies. However, when taking the correlated nature of the electron-hole pair into account, we show that such an increase must in fact be expected. Our work provides new microscopic insights into the ultrafast hole-transfer mechanism, and, more generally, highlights the potential of momentum microscopy to probe optically inaccessible correlated excitonic and electronic states of matter.


\vspace{1cm}
\noindent\textbf{Energy landscape and photoemission fingerprints of bright and dark excitons}

\begin{figure}[bt!]
    \centering
    \includegraphics[width=.9\linewidth]{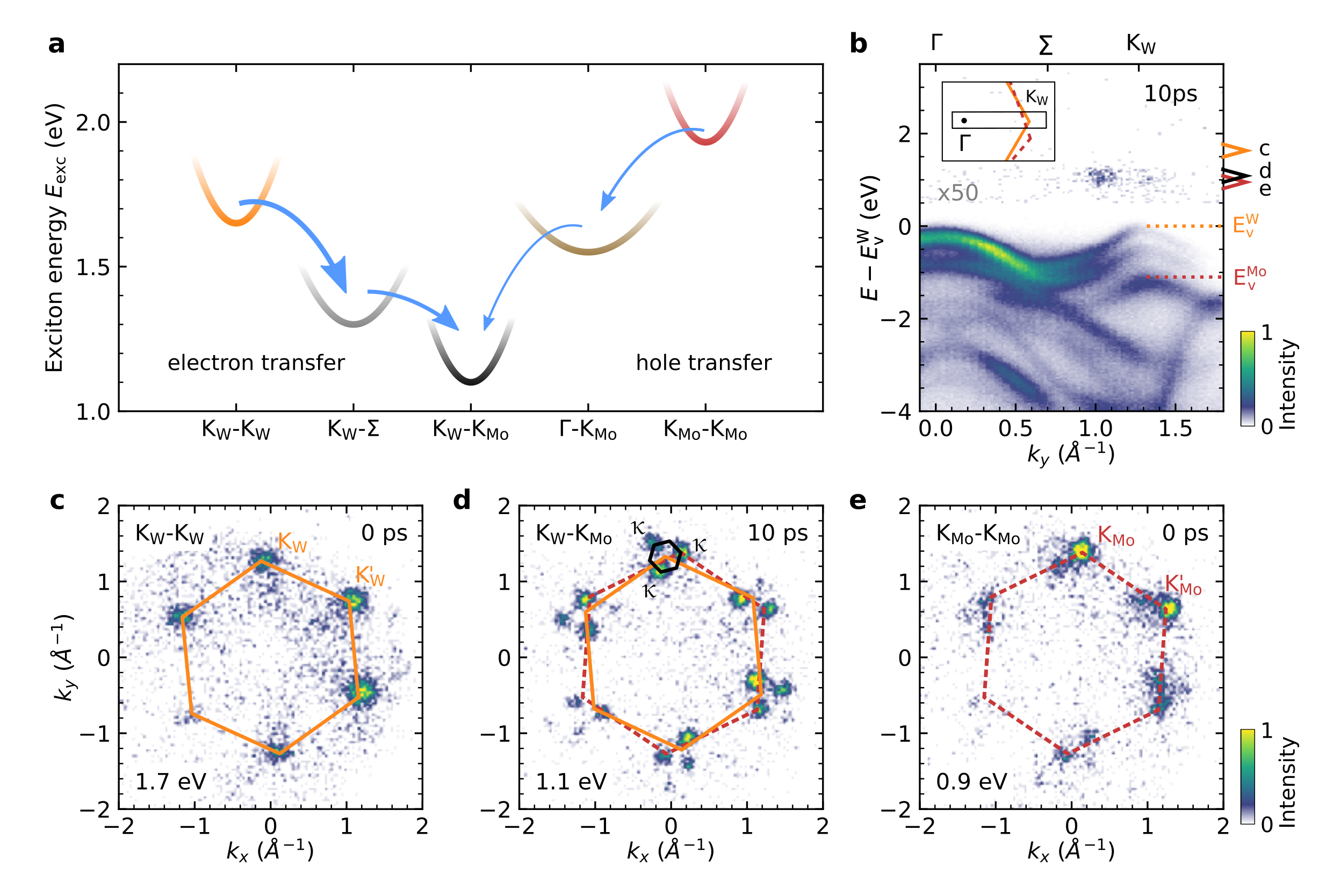}
    \caption{\textbf{Energy landscape and energy-momentum fingerprints of excitons in WSe$_2$/MoS$_2$.}
    (\textbf{a}) The calculated low energy exciton landscape is composed of bright intralayer K$_{\rm W}$-K$_{\rm W}$ (orange) and K$_{\rm Mo}$-K$_{\rm Mo}$ (dark red) excitons that can be resonantly excited with 1.7~eV and 1.9~eV light pulses, respectively. In a cascaded transition via layer hybridized K$_{\rm W}$-$\Sigma$ or $\Gamma$-K$_{\rm Mo}$ excitons, interlayer K$_{\rm W}$-K$_{\rm Mo}$ excitons (black) are formed. The calculated efficiency for each step in the exciton cascade is schematically encoded onto the width of the arrows.
    (\textbf{b}) Energy- and momentum-resolved photoemission spectrum along the $\Gamma$-$\Sigma$-K$_{\rm W}$ direction (inset) measured on the WSe$_2$/MoS$_2$ heterostructure after photoexcitation with 1.9~eV light pulses at a delay of 10~ps. The WSe$_2$ and MoS$_2$ valence band maxima are labelled with E$_{\rm v}^{\rm W}$ and E$_{\rm v}^{\rm Mo}$, respectively.
    (\textbf{c,d,e}) Photoemission momentum-fingerprints of the (\textbf{c}) intralayer K$_{\rm W}$-K$_{\rm W}$ exciton (0~ps), the (\textbf{d}) interlayer K$_{\rm W}$-K$_{\rm Mo}$ exciton (10~ps) and the (\textbf{e}) intralayer K$_{\rm Mo}$-K$_{\rm Mo}$ exciton (0~ps) after photoexcitation with 1.9~eV light pulses. The photoelectron energies of the momentum-maps are given in the figure with respect to the energy of the WSe$_2$ valence band maximum and indicated by colored arrowheads in (\textbf{b}). The Brillouin zones of WSe$_2$, MoS$_2$, and the moiré superlattice are overlaid on the data by orange, dark red (dashed), and black hexagons, respectively. 
    }
\end{figure}

We start the analysis of the hole-transfer dynamics by first calculating the full energy landscape of bright and dark excitons in the twisted WSe$_2$/MoS$_2$ heterostructure on a microscopic footing (details in supplementary text). The optically excited excitons and their cascaded relaxation via layer-hybridized intermediate states to the lowest energy excitons are illustrated in Fig.~2a. If the heterostructure is excited resonantly to the A1s-exciton of WSe$_2$ with 1.7~eV pulses, only intralayer K$_{\rm W}$-K$_{\rm W}$ A1s-excitons are optically excited and the formation of interlayer K$_{\rm W}$-K$_{\rm Mo}$ excitons via the electron-transfer process can be studied in detail~\cite{Schmitt22nat} (cf. Fig.~1b in the single particle picture and Fig. 2a, left side, in the exciton picture). In contrast, when the complementary process of hole-transfer across the WSe$_2$/MoS$_2$ interface is of interest, the dynamics must be initiated by an excitation with 1.9~eV light pulses in order to resonantly excite intralayer K$_{\rm Mo}$-K$_{\rm Mo}$ excitons (cf. Fig.~1c and Fig. 2a, right side). However, this also leads to an off-resonant excitation of K$_{\rm W}$-K$_{\rm W}$ excitons in WSe$_2$.

In order to differentiate the spectral contributions of different excitons, we apply our setup for femtosecond momentum microscopy, which provides direct access to the photoemission energy-momentum fingerprint of each exciton (Fig.~2b-e). This enables us to selectively follow the bright intralayer (K$_{\rm W}$-K$_{\rm W}$ and K$_{\rm Mo}$-K$_{\rm Mo}$, Fig.~2c,d) and dark interlayer (K$_{\rm W}$-K$_{\rm Mo}$, Fig.~2e) exciton formation and relaxation dynamics in a 9.8$\pm$0.8$^\circ$ twisted WSe$_2$/MoS$_2$ heterostructure (details in materials and methods and refs.~\cite{Keunecke20timeresolved,Keunecke20prb} ). In Fig.~2c, the momentum map of the intralayer K$_{\rm W}$-K$_{\rm W}$ exciton is shown, where the photoelectron is detected at the in-plane momenta of the K$_{\rm W}$ and K'$_{\rm W}$ valleys (0~ps). For better visibility, the Brillouin zone of WSe$_2$ is overlaid in orange. Complementary, in Fig.~2e, the momentum map of the K$_{\rm Mo}$-K$_{\rm Mo}$ exciton is shown, where the photoelectrons are found in the K$_{\rm Mo}$ and K'$_{\rm Mo}$ valleys of MoS$_2$ (cf. dark red hexagon, 0~ps). Note that the Brillouin zone of MoS$_2$ is rotated by 9.8$\pm$0.8$^\circ$ with respect to WSe$_2$. At a pump-probe delay of 10~ps, the major part of the intralayer excitons has decayed either via the electron- or the hole-transfer process, and spectral yield is dominated by the energetically most stable excitation, i.e., the interlayer K$_{\rm W}$-K$_{\rm Mo}$ excitons (fig.~S3). For these interlayer excitons, the electron- and the hole-contribution are now separated between both monolayers of the heterostructure, and the exciton photoemission momentum fingerprint has to be described within the moiré mini-Brillouin zones (mBz) built up by the $\kappa$ valleys whose in-plane momentum can be constructed by the reciprocal lattice vectors of WSe$_2$ and MoS$_2$ (Fig.~2d, black hexagon)~\cite{Schmitt22nat}. 

\vspace{.5cm}
\noindent\textbf{Hole- and electron-transfer dynamics}

\begin{figure}[bt!]
    \centering
    \includegraphics[width=.45\linewidth]{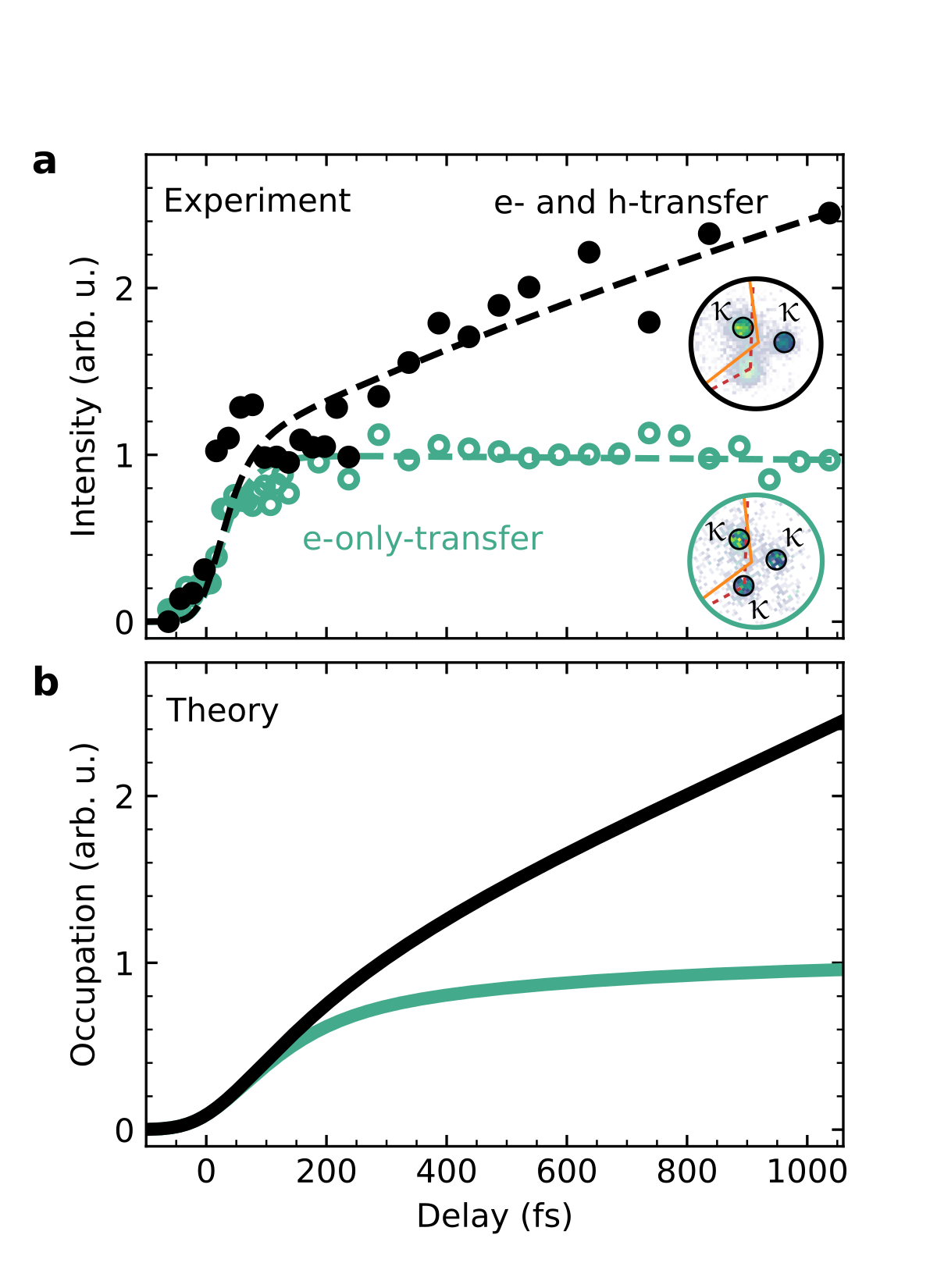}
    \caption{\textbf{Femto- to picosecond evolution of the hole- and electron-transfer dynamics.}
    (\textbf{a}) Direct comparison of the interlayer K$_{\rm W}$-K$_{\rm Mo}$ exciton formation dynamics if the heterostructure is excited resonant to intralayer K$_{\rm W}$-K$_{\rm W}$ exciton of WSe$_2$ (1.7~eV, green circles) or the intralayer K$_{\rm Mo}$-K$_{\rm Mo}$ exciton of MoS$_2$ (1.9~eV, black circles). While the electron-only-transfer process (1.7~eV) leads to a saturation of photoemission yield from interlayer K$_{\rm W}$-K$_{\rm Mo}$ excitons on the $<200$~fs timescale, the combined electron- and hole-transfer dynamics (1.9~eV) leads to an increasing photoemission yield beyond the 1~ps timescale.
    The insets show the momentum-filtered regions-of-interest (circles) used in the 1.7~eV (green contour) and 1.9~eV (black contour) measurement. The $\kappa$ valley that overlaps with the original K$_{\rm Mo}$ valley is excluded in the 1.9~eV measurement.
    (\textbf{b}) Microscopic model calculations of the interlayer K$_{\rm Mo}$-K$_{\rm W}$ exciton formation dynamics. The green curve describes the temporal evolution of the occupation of interlayer K$_{\rm W}$-K$_{\rm Mo}$ excitons after photoexcitation of intralayer K$_{\rm W}$-K$_{\rm W}$ excitons. For the black curve, the interlayer K$_{\rm W}$-K$_{\rm Mo}$ exciton formation dynamics is induced by the initial excitation of intralayer K$_{\rm W}$-K$_{\rm W}$ and K$_{\rm Mo}$-K$_{\rm Mo}$ excitons in the same occupation ratio as found in experiment (i.e., 1:5).
    }
\end{figure}

Having identified the exciton fingerprints in the photoemission experiment, we can now proceed with the analysis of the hole-transfer process. We first need to discern the contribution of non-resonantly excited K$_{\rm W}$-K$_{\rm W}$ excitons and their subsequent contribution to the formation dynamics of interlayer K$_{\rm W}$-K$_{\rm Mo}$ excitons via electron-transfer (see Fig.~1b). In order to do so, we make a direct comparison of interlayer K$_{\rm W}$-K$_{\rm Mo}$ exciton formation dynamics for 1.7~eV and 1.9~eV pumping. In Fig.~3a, the black data points show the pump-probe delay-dependent build-up of interlayer K$_{\rm W}$-K$_{\rm Mo}$ exciton density that is formed by electron-transfer \textit{and} hole-transfer processes (1.9~eV pump photons, exciton densities in WSe$_2$ and MoS$_2$: 7$\times 10^{11}$~cm$^{-2}$ and 3.5$\times 10^{12}$~cm$^{-2}$, respectively~\cite{Li14prb}). For comparison, the green data points show the pump-probe delay-dependent build-up of the interlayer K$_{\rm W}$-K$_{\rm Mo}$ exciton density that is formed \textit{only} via the electron transfer process (1.7~eV pump photons, exciton density: 5.4$\times 10^{12}$~cm$^{-2}$). It is directly obvious that there is a strong hierarchy of timescales for the electron- and hole-transfer processes: When considering the electron-only transfer process (green symbols), the interlayer exciton signal increases rapidly with pump-probe delay and saturates on the sub-200~fs timescale. A quantitative evaluation with rate equation modelling yields a formation time of $t_{\rm e-transfer}=$35$\pm$10~fs (cf. supplementary text). In contrast, the joint build-up of interlayer K$_{\rm W}$-K$_{\rm Mo}$ excitons via electron- and hole-transfer processes after 1.9~eV excitation increases beyond 1~ps (black symbols). In the global rate equation fit of both datasets, for a shared electron-transfer time $t_{\rm e-transfer}$, we then extract $t_{\rm h-transfer}=$2.9$\pm$0.6~ps for the hole-transfer process.

Hence, our experimental data implies that the interlayer hole-transfer mechanism across the WSe$_2$/MoS$_2$ heterointerface is significantly slower compared to the electron-transfer mechanism. In order to understand our findings, we carry out microscopic model calculations that incorporate exciton-light and exciton-phonon interactions (Fig.~3b, details in supplementary text and ref.~\cite{Meneghini22naturalsciences}). Clearly, we find an excellent qualitative agreement for the distinct formation timescales: The electron-only-transfer process saturates for delays $<200$~fs, while the combined electron- and hole-transfer dynamics lead to an increasing interlayer K$_{\rm W}$-K$_{\rm Mo}$ exciton density for delays beyond 1~ps. From the microscopic model calculations, we also find that the overall formation rate is dependent on the level of hybridization of the intermediate hybrid K$_{\rm W}$-$\Sigma$ and $\Gamma$-K$_{\rm Mo}$ excitons (fig.~S5) and the energetic offset between the different energy levels in the exciton cascade (Fig.~2a). Most importantly, in agreement with our experimental data, the calculations show that the electron-transfer process is always faster than the hole-transfer process for this material system.


\vspace{.5cm}
\noindent\textbf{The spectroscopic signature of a correlated hole-transfer process}

Based on this hierarchy of timescales between the electron- and the hole-transfer process, it is possible to separate the interlayer exciton formation dynamics: For delays $>200$~fs, the change in the exciton photoemission yield from interlayer K$_{\rm W}$-K$_{\rm Mo}$ exciton is mainly caused by hole-transfer processes. Hence, the final ambition of our work is the unambiguous discrimination of the photoemission spectral signature of intralayer K$_{\rm Mo}$-K$_{\rm Mo}$ and interlayer K$_{\rm W}$-K$_{\rm Mo}$ excitons, where, in both cases, the electron contribution to the exciton is situated in the conduction band minimum of the MoS$_2$ layer (cf. Fig.~1c). 

In the most naive picture of photoemission, it might be expected that trARPES only yields information on the exciton's electron. Hence, the experiment would not distinguish between photoelectrons being emitted from the conduction band minimum of MoS$_2$, irrespective whether they result from the break-up of intralayer K$_{\rm Mo}$-K$_{\rm Mo}$ or interlayer K$_{\rm W}$-K$_{\rm Mo}$ excitons (Fig.~1c). However, it is known that the spectral function in photoemission contains information about many-body-interactions~\cite{Damascelli03rmp}, and this is also the case for the correlated electron-hole pair. Importantly, this leads to a very non-intuitive and intriguing experimental observation.

\begin{figure}[b!]
    \centering
    \includegraphics[width=.9\linewidth]{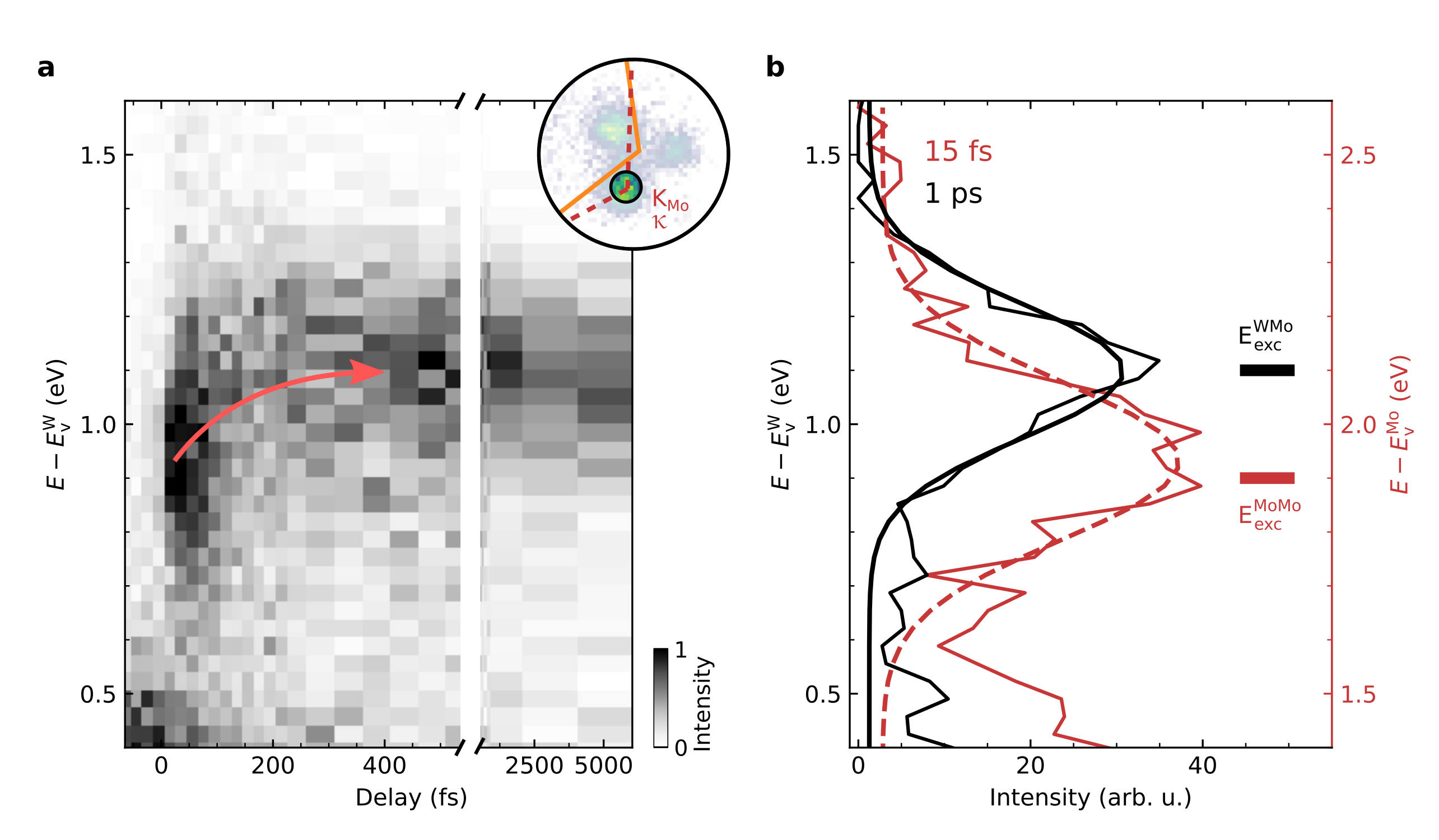}
    \caption{\textbf{Coulomb-correlation-induced excitonic energy fingerprints.}
    (\textbf{a}) Pump-probe delay evolution of the energy-distribution-curves (EDCs) filtered at the momentum-region of the K$_{\rm Mo}$ valley of MoS$_2$ (region-of-interest indicated in inset, 1.9~eV excitation). At this high-symmetry point, photoemission yield from intralayer K$_{\rm Mo}$-K$_{\rm Mo}$ and interlayer K$_{\rm W}$-K$_{\rm Mo}$ excitons is expected (cf. Fig.~1c). As intralayer K$_{\rm Mo}$-K$_{\rm Mo}$ excitons decay and form interlayer K$_{\rm W}$-K$_{\rm Mo}$ excitons, the peak maxima of the photoelectron energy shows an upshift by $\Delta E=0.17\pm 0.40$~eV (curved arrow). 
    (\textbf{b}) Selected EDCs for pump-probe delays of 15~fs (dark red) and 1~ps (black) illustrating an energetic upshift of the exciton photoemission signal. The horizontal bars indicate expected photoelectron energies for the intralayer K$_{\rm Mo}$-K$_{\rm Mo}$ (dark red) and interlayer K$_{\rm W}$-K$_{\rm Mo}$ (black) excitons calculated with Eq.~\ref{eq:energyconser} and data from photoluminescence measurements~\cite{Kunstmann18natphys,Karni19prl}. The left and right energy axis in black and dark red show the corresponding energy scales with respect to the valence band maximum of WSe$_2$ and MoS$_2$.
    }
\end{figure}

Figure~4a shows the pump-probe delay evolution of energy-distribution-curves (EDC) filtered for photoelectron yield at the $\kappa$ valley, whose momentum coincides with the K$_{\rm Mo}$ valley, i.e., the momentum region where photoelectron yield from intralayer K$_{\rm Mo}$-K$_{\rm Mo}$ and interlayer K$_{\rm W}$-K$_{\rm Mo}$ excitons is expected (Fig.~4a, inset). Astonishingly, we find that the mean energy of the photoelectrons shifts up as a function of pump-probe delay from E-E$_{\rm v}^{\rm W}=0.93\pm 0.03$~eV at 15~fs to E-E$_{\rm v}^{\rm W}=1.10\pm 0.03$~eV at 1~ps, i.e., a shift of $\Delta E =0.17\pm 0.04$~eV (Fig.~4b). At first glance, this is a surprising observation: In temporal overlap of the pump- and the probe laser pulses, the optical excitation deposits energy into the system, and the system subsequently relaxes from its excited state to energetically more favorable states via scattering processes. In consequence, energy-resolved pump-probe photoemission spectroscopies of single-particle charge carriers typically show that the mean kinetic energy of the photoelectrons decreases with pump-probe delay~\cite{Bauer15pss}. An increasing mean kinetic energy might indicate higher-order scattering processes such as Auger recombination~\cite{Mathias16natcom,Erkensten21Auger}. For Auger recombination, however, we would expect to observe a decreasing mean kinetic energy on the few-picosecond timescale as the overall exciton density and thus the efficiency for Auger recombination decreases. However, the long-time evaluation of the mean photoelectron energy clearly excludes this scenario (Fig.~4a). In addition, by evaluating the pump-probe delay evolution of the energy position of the MoS$_2$ valence band maxima, we can exclude a photo-induced renormalization of the band energies~\cite{Chernikov15natpho,Liu19prl} (fig.~S4). We thus search for the origin of the apparent increase of the mean kinetic energy beyond the single-particle picture, i.e., in the photoemission from excitons whose occupation is dynamically transferring from intralayer K$_{\rm Mo}$-K$_{\rm Mo}$ to interlayer K$_{\rm W}$-K$_{\rm Mo}$ excitons.



So far, we have referenced the energies of all emitted single-particle photoelectrons to the valence band maximum of WSe$_2$ (left energy axis in Fig.~4b). However, especially for the intralayer K$_{\rm Mo}$-K$_{\rm Mo}$ exciton that fully resides in the MoS$_2$ layer, this is clearly not the intrinsically relevant energy axis. We overcome this shortcoming by using an energy scale that is more direct to photoemission from excitons by relating the total energy before ($E=E_0+E_{\rm exc}+\hbar\omega$) and after ($E=E_0-E_{\rm hole}+E_{\rm elec}$) the break-up of the correlated electron-hole pair~\cite{Weinelt04prl} (with $E_{\rm hole}$ denoting the energy of the single-particle hole-state, $E_{\rm elec}$  the energy of the single-particle electron-state, $E_0$  the ground state energy, $E_{\rm exc}$ the exciton energy, and $\hbar\omega$ denoting the photon energy). As energy needs to be conserved when the exciton is broken, the energy of the detected single-particle electron can be expressed as
\begin{equation}
    E_{\rm elec}=E_{\rm hole}+E_{\rm exc}+\hbar\omega.
    \label{eq:energyconser}
\end{equation}
Importantly, Eq.~\ref{eq:energyconser} sets the energy of the single-particle hole $E_{\rm hole}$ remaining in the sample as a natural reference point of the photoelectron energy axis for each probed exciton. For the intralayer K$_{\rm Mo}$-K$_{\rm Mo}$ excitons and the interlayer K$_{\rm W}$-K$_{\rm Mo}$ excitons, respectively, the valence band maxima of MoS$_2$ (E$_{\rm v}^{\rm Mo}$) and WSe$_2$ (E$_{\rm v}^{\rm W}$) set the energy scale (energies labelled in Fig.~2b). Using the respective exciton energies reported from photoluminescence spectroscopy (E$_{\rm exc}^{\rm MoMo}=1.9$~eV and E$_{\rm exc}^{\rm WMo}=1.1$~eV)~\cite{Karni19prl,Kunstmann18natphys}, hence, we expect photoemission yield from the intralayer K$_{\rm Mo}$-K$_{\rm Mo}$ excitons and the interlayer K$_{\rm W}$-K$_{\rm Mo}$ excitons at E-E$_{\rm v}^{\rm Mo}=1.9$~eV and E-E$_{\rm v}^{\rm W}=1.1$~eV, respectively (horizontal lines in Fig.~4B; left-hand and right-hand side energy axis in Fig.~4b). Indeed, the calculated energies perfectly agree with the measured photoelectron energy distributions. We thus explain the upshift of the mean photoelectron energy as a direct consequence of the correlated nature of the electron-hole pairs: Despite energy relaxation, the single-particle photoelectron is found at an increased photoemission energy after the exciton's hole-transfer across the TMD interface.

\vspace{.5cm}
\noindent\textbf{Conclusion}

In summary, we have shown that femtosecond momentum microscopy is a powerful tool to study correlations and cooperative exciton behavior in moiré heterostructures. Exemplary, we show that the photoelectron of the correlated two-particle exciton contains direct information about the hole state. We use this correlation in combination with microscopic and material-specific theory to directly follow an ultrafast interlayer hole transfer process that would otherwise be elusive. Our work opens up new means for the future study of correlated states of matter in two-dimensional quantum materials.


\section{ACKNOWLEDGEMENTS}

This work was funded by the Deutsche Forschungsgemeinschaft (DFG, German Research Foundation) - 432680300/SFB 1456, project B01, 217133147/SFB 1073, projects B07 and B10, and 223848855/SFB 1083, project B9. A.A. and S.H. acknowledge funding from EPSRC (EP/T001038/1, EP/P005152/1). A.A. acknowledges financial support by the Saudi Arabian Ministry of Higher Education. E. M. acknowledges support from the European Union's Horizon 2020 research and innovation programme under grant agreement no. 881603 (Graphene Flagship). K.W. and T.T. acknowledge support from JSPS KAKENHI (Grant Numbers 19H05790, 20H00354 and 21H05233).

\section{AUTHOR CONTRIBUTIONS}
D.St., S.S., R.T.W., G.S.M.J., S.H., S.B., E.M., M.R. and S.M. conceived the research. D.Sch., J.P.B. and W.B. carried out the time-resolved momentum microscopy experiments. J.P.B. and D.Schm. analyzed the data. G.M. performed the microscopic model calculations. A.A. fabricated the heterostructure sample. All authors discussed the results. M.R. and S.M. were responsible for the overall project direction and wrote the manuscript with contributions from all co-authors. K.W. and T.T. synthesized the hBN crystals.


\clearpage\newpage\newpage

\noindent \textbf{Supplementary Materials to\\ Probing correlations in the exciton landscape of a moiré heterostructure}

\renewcommand{\figurename}{Figure}
\renewcommand{\thefigure}{S\arabic{figure}}
\def\bibsection{\section*{\refname}} 
\setcounter{figure}{0}
\renewcommand{\theequation}{S\arabic{equation}}
\setcounter{equation}{0}


\section{Materials and Methods}

The time- and angle-resolved photoemission data is measured with a time-of-flight momentum microscope (Surface Concept, ToF-MM)~\cite{medjanik_direct_2017,kromker_development_2008} that is connected to a table-top high harmonic generation (HHG) beamline driven by a 300~W fiber laser system (AFS Jena)~\cite{Keunecke20prb,Duvel22nanolett}. The overall experimental setup and its application to exfoliated two-dimensional materials is described in refs.~\cite{Keunecke20timeresolved} and~\cite{Schmitt22nat}, respectively.

In all experiments, the exciton dynamics are induced by resonant optical excitation of the A-excitons of WSe$_2$ or MoS$_2$. Therefore, 1.7~eV and 1.9~eV pump pulses with a duration of 50~fs are used ($s$-polarized), respectively. After a variable pump-probe delay, photoemission is induced by 26.5~eV light pulses (20~fs, $p$-polarized).

The 9.8$\pm$0.8$^\circ$ twisted WSe$_2$/MoS$_2$ heterostructure is stamped onto a 20-30~nm thick hBN~\cite{Taniguchi07jcg} spacer layer and a p$^+$-doped native oxide silicon waver. Prior to the momentum microscopy experiments, the sample is annealed for 1~h to 670~K. Details on the sample fabrication and characterization (e.g., twist angle) are described in ref.~\cite{Schmitt22nat}.

\clearpage
\section{Correction of rigid band shifts}

\begin{figure}[hbt!]
    \centering
    \includegraphics[width=.95\linewidth]{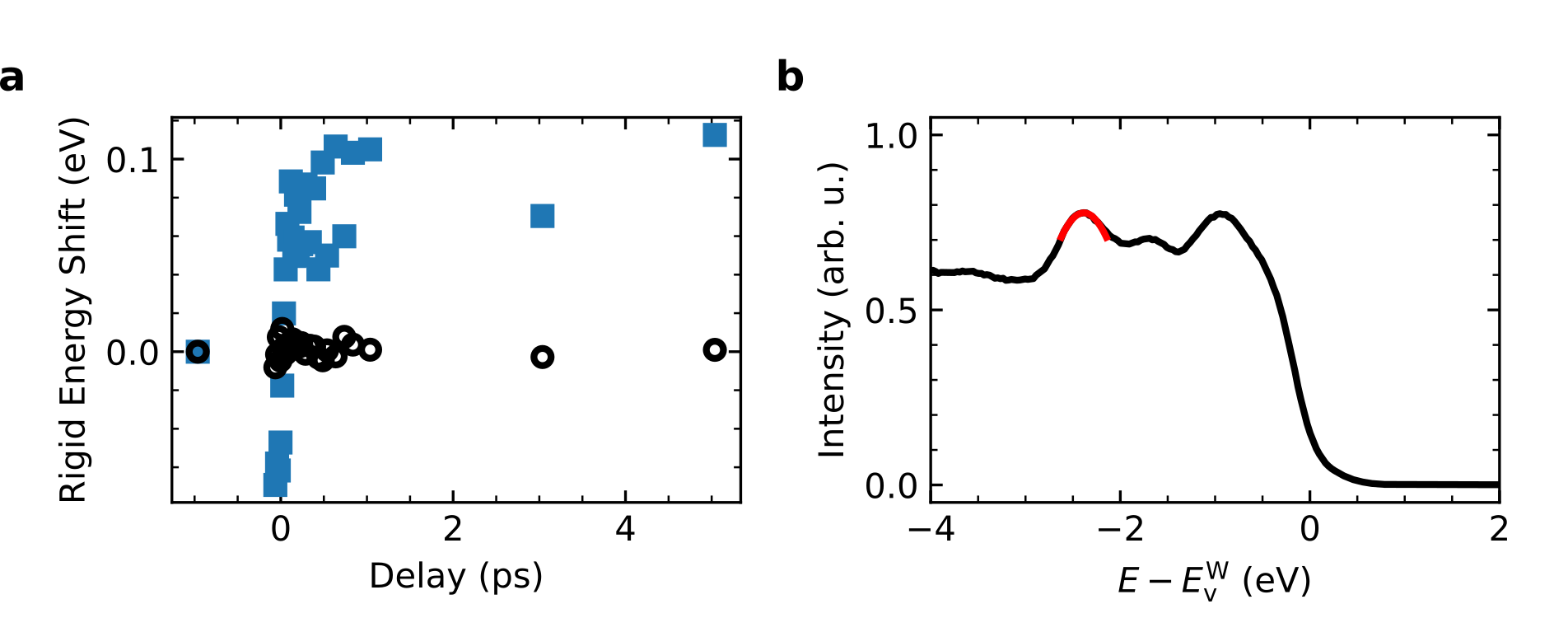}
    \caption{\textbf{Correction of rigid energy shifts.}
    (\textbf{a}) At each pump-probe delay, the momentum-integrated energy spectrum has a distinct rigid energy shift (blue squares) that is evaluated by fitting the red marked peak close to E-E$_v^W$= -2.4~eV in (\textbf{b}). After the correction of this rigid energy shift, all energy-distribution-curves are aligned (black circle data points).}
    \label{rigid}
\end{figure}

As a result of pump- and probe-induced space-charge and surface photovoltage effects, we observe transient energy shifts of the momentum-integrated photoemission spectrum by $\pm$80~meV~\cite{Schonhense21rsi} (Fig.~\ref{rigid}). This energy offset is extracted from the data by fitting the spectral weight maxima at $\approx$-2.4~eV (Fig. S1b) for each pump-probe delay. The blue and the black data points show the pump-probe dependence of this peak before and after correction, respectively. The correction is done prior to the data analysis discussed in the main text.

\clearpage

\section{Quantitative analysis of the exciton dynamics - Rate equation modelling}

\begin{figure}[hbt]
    \centering
    \includegraphics[width=.95\linewidth]{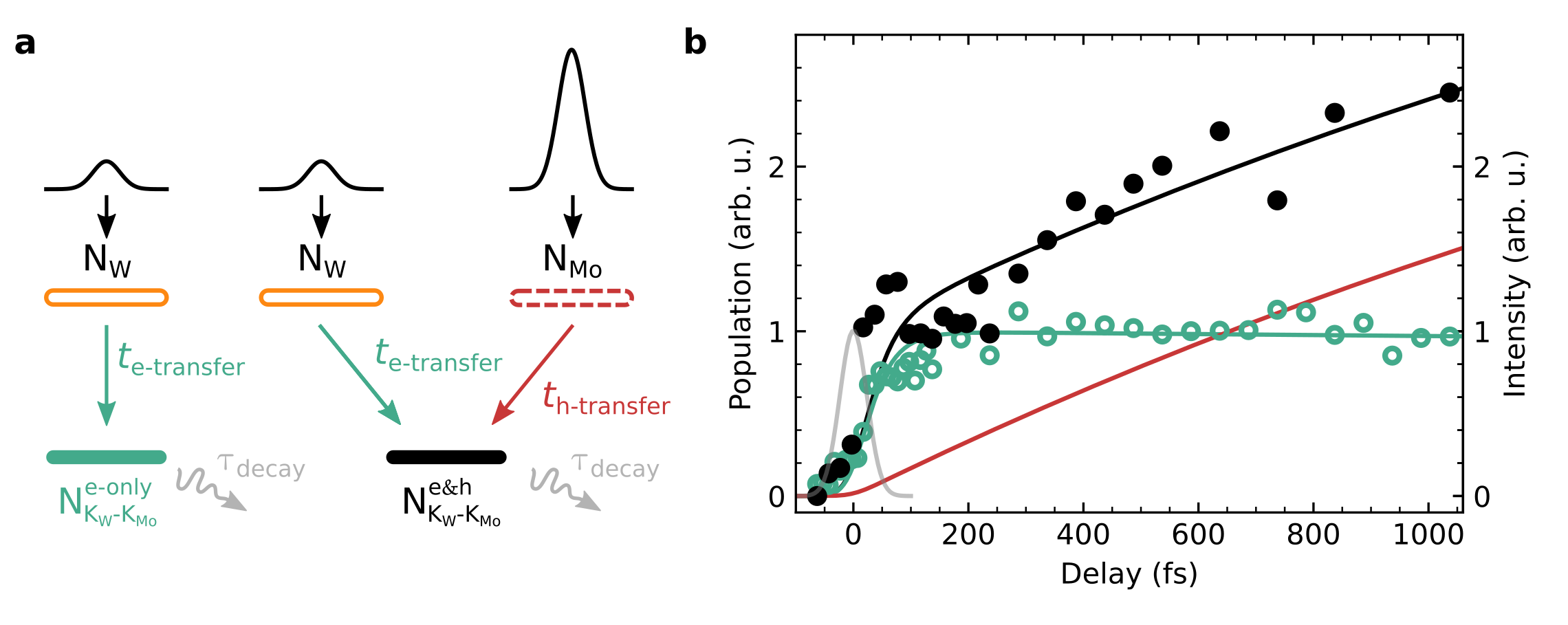}
    \caption{\textbf{Quantitative analysis of the interlayer K$_{\rm W}$-K$_{\rm Mo}$ exciton formation dynamics.}
    (\textbf{a}) Schematic overview of the rate equation model: Exciton population in the WSe$_2$ (MoS$_2$) layer $N_{\rm W}$ ($N_{\rm Mo}$) is excited by a Gaussian pump pulse. Electron (hole) charge-transfer leads to the formation of interlayer K$_{\rm W}$-K$_{\rm Mo}$ excitons ($N^{\rm e \& h}_{\rm K_W-K_{Mo}}$) with the scattering time $t_{\rm e-transfer}$ ($t_{\rm h-transfer}$). $N^{\rm e-only}_{\rm K_W-K_{Mo}}$ describes the case for the electron-only charge-transfer process. 
    (\textbf{b}) The pump-probe delay-dependent build-up of photoemission intensity of interlayer K$_{\rm W}$-K$_{\rm Mo}$ excitons is shown after resonant excitation of K$_{\rm Mo}$-K$_{\rm Mo}$ excitons in MoS$_2$ (1.9~eV, black circles) and after resonant excitation of K$_{\rm W}$-K$_{\rm W}$ excitons in WSe$_2$ (1.7~eV, green circles), respectively. 
    Note that resonant excitation of K$_{\rm Mo}$-K$_{\rm Mo}$ excitons also leads to off-resonant excitation of K$_{\rm W}$-K$_{\rm W}$ excitons, so that the interlayer exciton is build-up both by hole and electron transfer.
    Solid lines depict best fit results of the rate equation model. Green and dark red solid lines describe the proportion of the interlayer exciton population that is created due to the electron-only ($N^{\rm e-only}_{\rm K_W-K_{Mo}}$) and hole-only charge transfer. The black line corresponds to the sum of electron- and hole-transfer processes $N^{\rm e \& h}_{\rm K_W-K_{Mo}}$.}
    \label{rateequation}
\end{figure}

Fig. 3a in the main text and Fig.~\ref{rateequation}b show the pump-probe delay-dependent photoemission intensity from interlayer K$_{\rm W}$-K$_{\rm Mo}$ excitons in the case that electron- and hole-transfer processes (black circles, 1.9~eV) or electron-only-transfer  processes (green circles, 1.7~eV) contribute. In order to quantitatively analyze the characteristic formation dynamics, we apply a rate equation model to fit the experimental data. The model is schematically shown in Fig.~\ref{rateequation}a and the rate equations are listed in the following

\begin{align}
   \frac{d N_{\rm W}}{d t} &= g_{\rm W}(t), \\
   \frac{d N_{\rm Mo}}{d t} &= g_{\rm Mo}(t), \\
   \frac{d N^{\rm e \& h}_{\rm K_W-K_{Mo}}}{d t} &= \frac{N_{\rm Mo}}{t_{\rm h-transfer}} + \frac{N_{\rm W}}{t_{\rm e-transfer}} -\frac{N^{\rm e \& h}_{\rm K_W-K_{Mo}}}{\tau_{\rm decay}},\\
   \frac{d N^{\rm e-only}_{\rm K_W-K_{Mo}}}{d t} &=  \frac{N_{\rm W}}{t_{\rm e-transfer}} - \frac{N^{\rm e-only}_{\rm K_W-K_{Mo}}}{\tau_{\rm decay}}.
\end{align}

\noindent $N_{\rm W}$ and $N_{\rm Mo}$ are the intralayer exciton occupation in the WSe$_2$ and MoS$_2$ layer, respectively. A Gaussian-shaped excitation $g_{\rm W}(t)$ is used to initiate the dynamics with the experimental pump pulse duration of ${\rm FWHM} = 50$~fs. To account for the 5:1 ratio in the exciton densities in the 1.9~eV pump measurement, the MoS$_2$ layer $N_{\rm Mo}$ gets excited with $g_{\rm Mo}(t) = 5 \cdot g_{\rm W}(t)$. For the 1.9~eV pumped data, $t_{\rm e-transfer}$ describes the scattering time from $N_{\rm W}$ to the interlayer exciton state $N^{\rm e \& h}_{\rm K_W-K_{Mo}}$, i.e. the electron charge-transfer time. In addition, $t_{\rm h-transfer}$ is the scattering time from $N_{\rm Mo}$ into the $N^{\rm e \& h}_{\rm K_W-K_{Mo}}$ state, i.e. the hole charge-transfer time. Moreover, the interlayer exciton state depopulates with a fixed decay time of $\tau_{\rm decay} = 33$~ps, which was estimated from an exponential decay fit to the long-term dynamics~\cite{Schmitt22nat}. In a similar manner, for the 1.7~eV pumped data, $N^{\rm e-only}_{\rm K_W-K_{Mo}}$ describes the evolution of the interlayer exciton occupation if only electron-transfer-proccesses occur.

In the fit routine, equations (S1-S4) are solved numerically yielding the delay-dependent evolution of the states $N^{\rm e \& h}_{\rm K_W-K_{Mo}}$ and $N^{\rm e-only}_{\rm K_W-K_{Mo}}$. We apply a global fit approach in which, simultaneously, the occupation $N^{\rm e \& h}_{\rm K_W-K_{Mo}}$ is compared to the 1.9~eV pumped data and the occupation $N^{\rm e-only}_{\rm K_W-K_{Mo}}$ is compared to the 1.7~eV pumped data. Optimization for the shared fit parameters $t_{\rm e-transfer}$ and $t_{\rm h-transfer}$ results in $t_{\rm e-transfer} = 35\pm10{\rm~fs}$ and $t_{\rm h-transfer}  = 2.9\pm0.6{\rm~ps}$.

\clearpage


\section{Femtosecond dynamics of intra- and interlayer excitons}

\begin{figure}[hbt!]
    \centering
    \includegraphics[width=.75\linewidth]{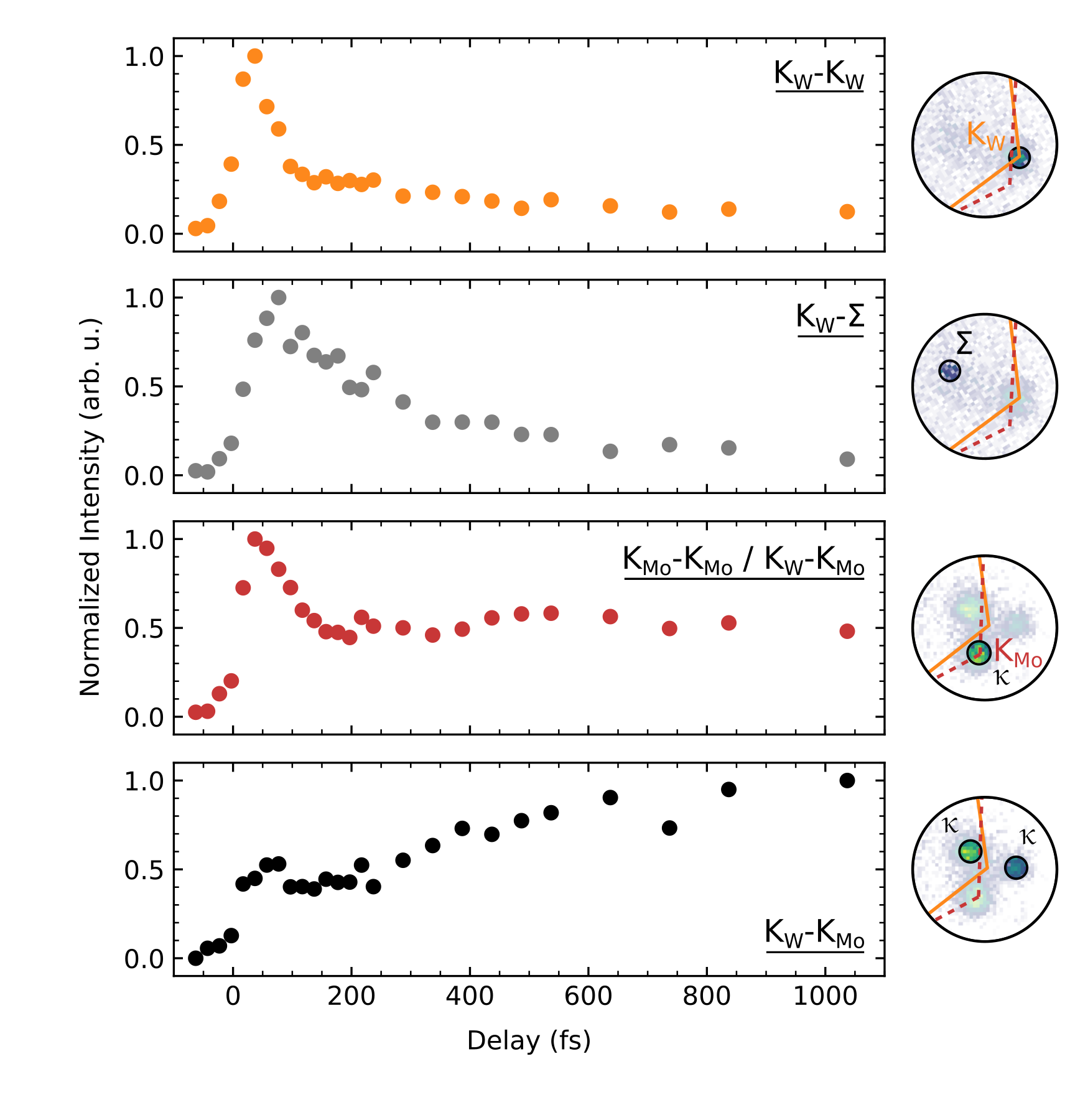}
    \caption{\textbf{Femtosecond intra- to interlayer exciton transfer dynamics.}
    The dynamics for the different exciton signals are depicted: K$_{\rm W}$-K$_{\rm W}$ (orange), K$_{\rm W}$-$\Sigma$ (grey), K$_{\rm Mo}$-K$_{\rm Mo}$ and K$_{\rm Mo}$-K$_{\rm W}$ (dark red) and K$_{\rm W}$-K$_{\rm Mo}$ (black). The round insets show the position of the momentum apertures used for filtering the exciton photoemission signatures. If the $\kappa$ valley coincides with the K$_{\rm Mo}$ valley (dark red), photoemission yield is composed of contributions from intralayer K$_{\rm Mo}$-K$_{\rm Mo}$ and interlayer K$_{\rm W}$-K$_{\rm W}$ excitons (cf. Fig~4 of the main text). In the case that those high-symmetry points do not overlap (black), only photoemission signal from interlayer K$_{\rm W}$-K$_{\rm Mo}$ excitons is detected (cf. Fig.~3 of the main text).}
    \label{dynamics-all}
\end{figure}

Figure~\ref{dynamics-all} shows an overview of the pump-probe delay-dependent evolution of photoemission intensity for all measured excitons after excitation with 1.9~eV light pulses. The first two rows show the optical excitation of intralayer K$_{\rm W}$-K$_{\rm W}$ excitons (orange) and the subsequent formation of hybrid K$_{\rm W}$-$\Sigma$ excitons (grey). The bottom two panels show the photoemission intensity of selected $\kappa$ valleys of the moiré mBz. In the case that the $\kappa$ valley overlaps with the K$_{\rm Mo}$ valley (dark red), photoemission intensity is composed of signal from intralayer K$_{\rm Mo}$-K$_{\rm Mo}$ and interlayer K$_{\rm Mo}$-K$_{\rm W}$ excitons. This $\kappa$ valley is evaluated in Fig.~4 of the main text. Complementary, if the two $\kappa$ valleys are evaluated (black) that do not overlap with the K$_{\rm Mo}$ valley, only photoemission signal from interlayer K$_{\rm Mo}$-K$_{\rm W}$ excitons is detected. These data is shown in Fig.~3a of the main text.

\clearpage


\section{Excluding photo-induced band renormalizations}

The major spectroscopic signature of interest in our manuscript is a pump-probe delay-dependent upshift of the energy of photoelectrons being emitted from excitons (Fig.~4). We attribute this energy upshift to the formation of interlayer K$_{\rm Mo}$-K$_{\rm W}$ excitons from intralayer K$_{\rm Mo}$-K$_{\rm Mo}$ excitons. However, it is well-known that photo-induced band renormalizations~\cite{Chernikov15natpho} can lead to a similar shift of photoemission signatures~\cite{Liu19prl}, which, hence, must be excluded.

\begin{figure}[hbt]
    \centering
    \includegraphics[width=.75\linewidth]{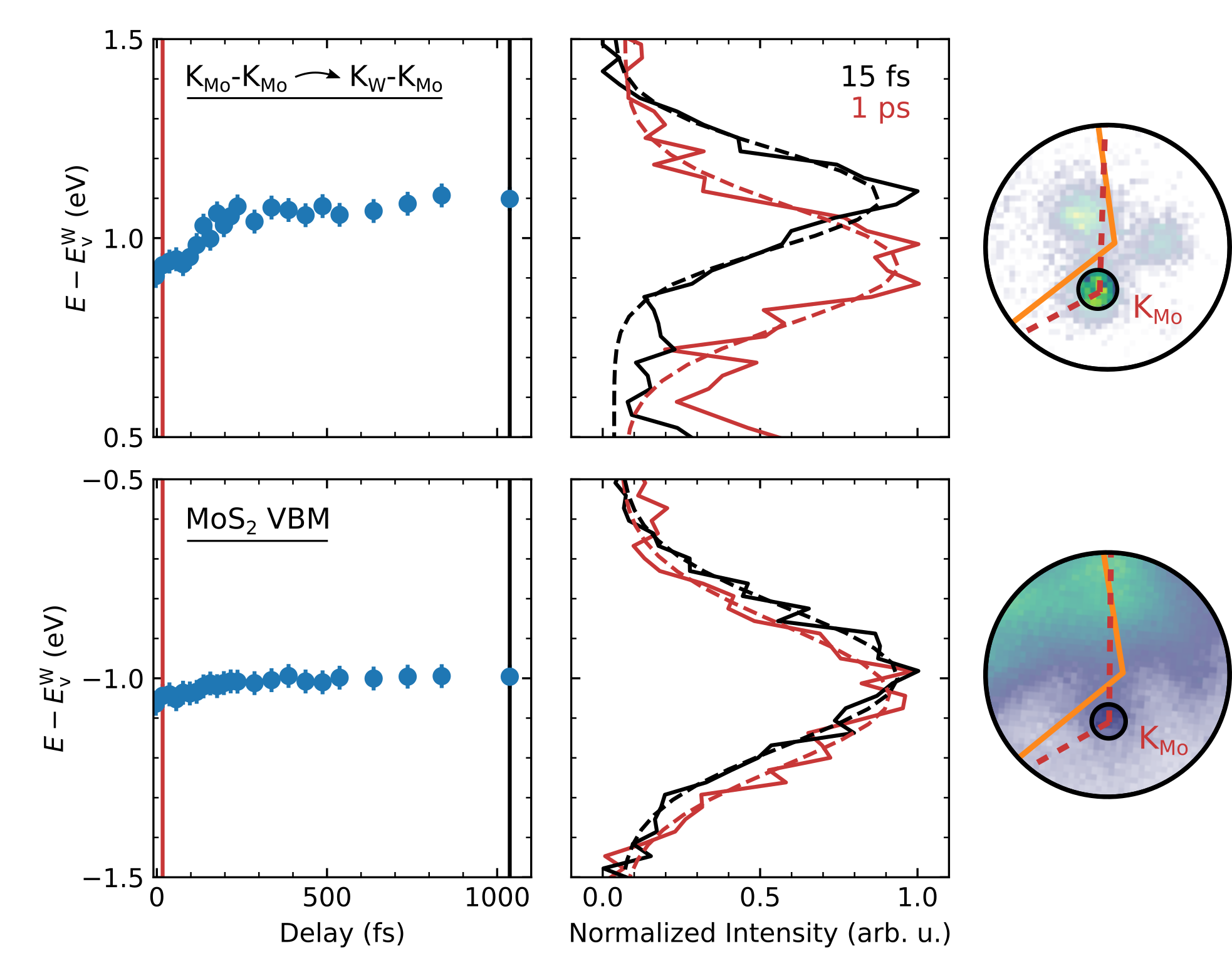}
    \caption{\textbf{Excluding photo-induced band renormalizations.}
    The top and the bottom rows show the peak position of the exciton photoemission signal and the MoS$_2$ valence band maximum, respectively. 
    In the middle panels, selected EDCs taken at the K$_{\rm Mo}$ valley are shown for 15~fs (dark red) and 1~ps (black). In the left panels, the fitted peak maxima of such EDCs are plotted as a function of pump-probe delay. The right panels show the filtered momentum regions (black circles), whereas the momentum-momentum maps are taken at the respective energies of the excitonic photoemission signal and the MoS$_2$ valence band maximum. }
    \label{excluding-photo-shift}
\end{figure}

In addition to photoemission signals from excitons, the multidimensional data acquisition scheme allows to monitor the energetic position of the occupied valence band of MoS$_2$. If the electronic bands would renormalize in response to the optical excitation, we would expect to observe an energetic shift of this occupied valence band~\cite{Liu19prl}. In Fig.~\ref{excluding-photo-shift}, we directly compare the energy position of the MoS$_2$ valence band and the excitonic photoemission signal. We observe that after the excitation with the pump pulse the excitonic peak position at the K$_{\rm Mo}$ point exhibits an upshift, while the valence band maximum of the MoS$_2$ layer remains comparably constant. Hence, we can exclude photo-induced band renormalizations as the origin for the energetic upshift of the main photoemission signal in Fig.~4. 

\clearpage


\section{Microscopic modelling} 
\begin{figure}[hbt!]
    \centering
    \includegraphics[width=.8\linewidth]{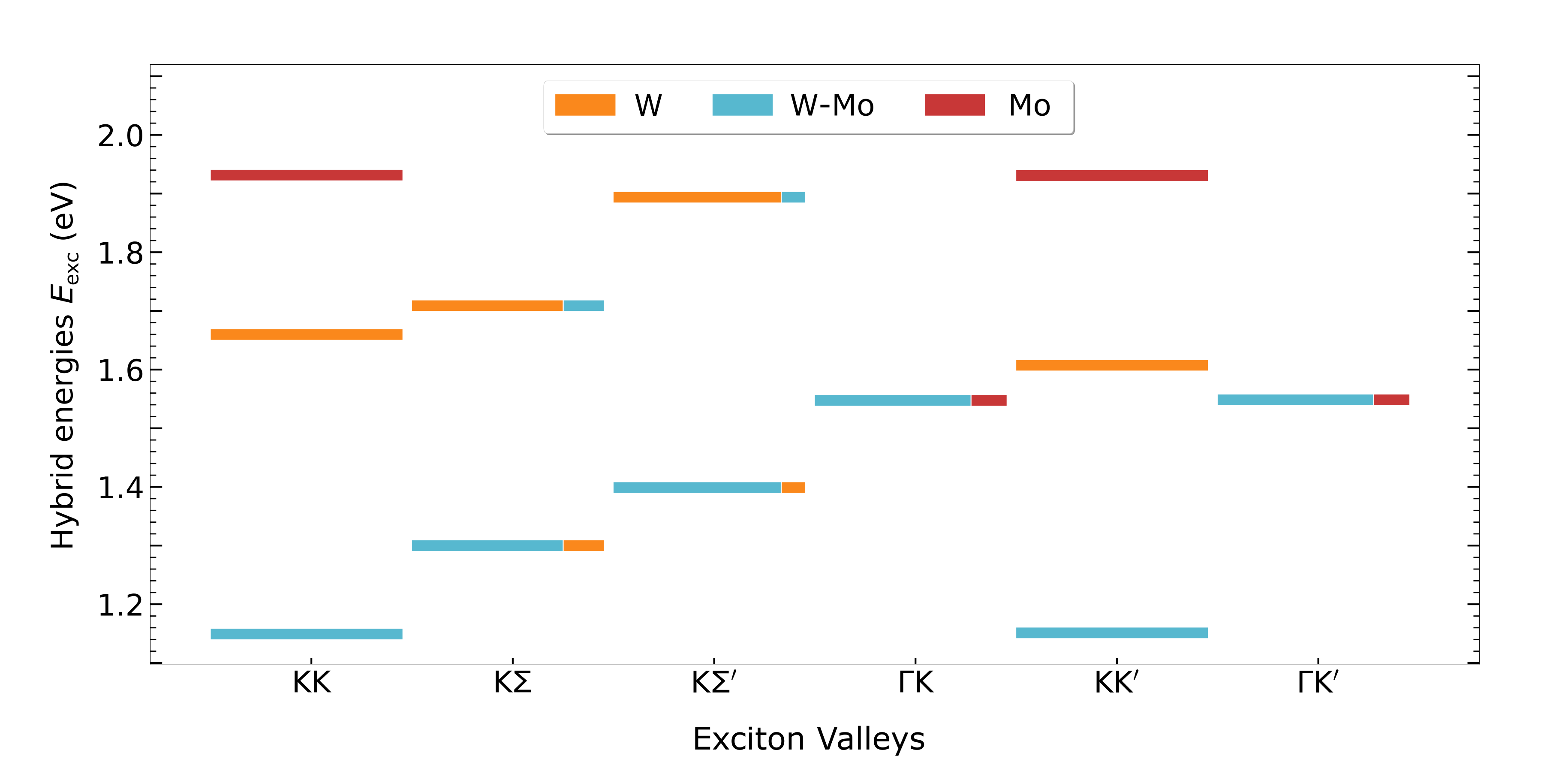}
    \caption{Hybrid-exciton energy landscape for the WSe$_2$-MoS$_2$ heterostructure.  We use different colors for depicting the percentage of intralayer tungsten (W, orange), intralayer molybdenum (Mo, red) or interlayer (blue) exciton character of the corresponding states. 
   Due to the strong tunneling experienced by electrons, K$\Sigma^{(\prime)}$ and $\Gamma$K states are strongly hybridized. Note that we plot only a selection of low-energy hybrid exciton states contributing directly to the relaxation dynamics.}
    \label{energy_landscape}
\end{figure}

In this section, we introduce the main concepts of the  theoretical approach applied to calculate the dynamics in TMD bilayers. We start with the excitonic Hamilton operator 
\begin{equation}\label{eq:4}
    H = H_0 + H_T = 
    \sum_{\mu, {\bf Q}}{ E^{\mu }_{{\bf Q}} X^{\mu\dagger}_{{\bf Q}} X^{\mu}_{{\bf Q}} }
    + \sum_{\substack{\mu,\nu,{\bf Q}}}{ \Tb_{\mu \nu}  {X^{\mu\dagger }_{{\bf Q}}}X^{\nu}_{{\bf Q}} }
\end{equation}
where we used the superindex $\mu = (n^\mu, \zeta^\mu_e,\zeta^\mu_h,L^\mu_e,L^\mu_h)$ to describe the exciton states, $E^\mu_{\bf Q} = E^c_{\zeta^\mu L^\mu_e}-E^v_{\zeta^\mu L^\mu_h} + E^\mu_{bind} +  E^\mu_{ {\bf Q},kin}$ are the excitonic energies, where $E^\mu_{bind}$ are obtained after solving a bilayer Wannier equation \cite{ovesen2019interlayer, brem2020hybridized}, $E^{c/v}_{\zeta^\mu L^\mu_e}$ are the conduction and valence band energy and $E^\mu_{{\bf Q},kin} = \hbar^2 {\bf Q}^2/ (2 M^\mu)$ is the kinetic energy of the exciton with mass $M^\mu =(m^\mu_e+m^\mu_h)$. Moreover we introduced the excitonic tunneling between the TMD monolayers with the tunnelling matrix elements
\begin{equation}\label{eq:5}
  \Tb_{\mu \nu} = (\delta_{L^{\mu}_h L^{\nu}_h} (1- \delta_{L^{\mu}_e L^{\nu}_e})  \delta_{\zeta^\mu \zeta^\nu} T^c_{\mu_e,\nu_e} - \delta_{L^{\mu}_e L^{\nu}_e} (1- \delta_{L^{\mu}_h L^{\nu}_h}) \delta_{\zeta^\mu \zeta^\nu} T^v_{\mu_h,\nu_h})  \sum_{\bf k}{ \psi^{\mu*}({\bf k} )\psi^{\nu}({\bf k}) },
\end{equation}
where $\psi^{\mu}$ is the excitonic wave function of the state $\mu$ defined over the relative momentum between electron and hole, $T^\lambda_{i j} = \bra{\lambda i {\bf p} } H \ket{\lambda j {\bf p}} (1-\delta_{L_i L_j}) \delta_{\zeta_i \zeta_j}$ are the electronic tunneling elements obtained by averaging DFT values of MoSe$_2$-WSe$_2$ and MoS$_2$-WS$_2$ heterostructures in \cite{PhysRevResearch.3.043217}. 
Diagonalizing Eq. \ref{eq:4} leads to a new set of hybrid excitonic energies $\Ea^\eta_{\bf Q}$ that are obtained by solving the hybrid eigenvalue equation \cite{brem2020hybridized, PhysRevResearch.3.043217},
\begin{equation}\label{eq:6}
    E^\mu_{\bf Q} c^\eta_\mu({\bf Q}) + \sum_{\nu}{ \Tb^{}_{\mu \nu} c^\eta_\nu({\bf Q}) } = \Ea^\eta_{\bf Q} c^\eta_\mu({\bf Q}).
\end{equation}
Now, we can define a diagonal hybrid exciton Hamiltonian \cite{Meneghini22naturalsciences, Schmitt22nat}
\begin{equation}\label{eq:7}
     H= \sum_{\eta}{\Ea^\eta_{\bf Q}  Y^{\eta \dagger}_{\bf Q}  Y^\eta_{\bf Q} } 
\end{equation} 
with the hybrid exciton annihilation/creation operators $Y^{\eta(\dagger)}_{\bf Q} = \sum_{ \mu }{ c^\eta_\mu({\bf Q}) X^{\mu(\dagger)}_{\bf Q}}$.
Evaluating the above eigenvalue equation, we predict the hybrid exciton energy landscape for the investigated  WSe$_2$-MoS$_2$ heterostructure, cf. Fig. \ref{energy_landscape}.

The hybrid exciton-phonon scattering plays a crucial role at the low excitation regime \cite{brem2018exciton, Meneghini22naturalsciences}.
The  corresponding Hamiltonian can be written as \cite{brem2020hybridized}
\begin{equation}\label{eq:9}
    H_{Y-ph} = \sum_{j,{\bf Q},{\bf q}, \eta,\xi}{ \tilde{\Da}^{\xi\eta}_{j,{\bf q},{\bf Q}} Y^{\xi\dagger }_{{\bf Q + q}}Y^{\eta}_{{\bf Q}} b_{j,{\bf q}} } + h.c.
\end{equation}
with the hybrid exciton-phonon coupling $\tilde{\Da}^{\xi\eta}_{j,{\bf q},{\bf Q}}$. The electron-phonon coupling matrix elements, single-particle energies and effective masses are taken from DFPT calculations \cite{PhysRevB.90.045422}.
The excitation of the system through a laser pulse is described semi-classically via the minimal-coupling Hamiltonian that can be written as \cite{brem2020hybridized}
\begin{equation}
    H_{Y-l} = \sum_{\sigma,{\bf Q},\eta}{ {\bf A} \cdot \tilde{\mathcal{M}}^\eta_{\sigma {\bf Q}} Y^\eta_{\bf Q_{\parallel} }} + h.c.\\
\end{equation}
with hybrid exciton-light coupling $\tilde{\mathcal{M}}^\eta_{\sigma {\bf Q}}$.
Details on the transformation and the definition of the hybrid interaction matrix elements and couplings are given in Ref. \cite{brem2020hybridized, PhysRevResearch.3.043217}.

Exploiting the Heisenberg equation of motion for the hybrid occupation $N^\eta = \langle Y^{\eta\dagger }Y^{\eta}  \rangle$, including $H = H_Y + H_{Y-ph}+ H_{Y-l}$, and truncating the Martin-Schwinger hierarchy using a second order Born-Markov approximation \cite{kira2006many,haug2009quantum,malic2013graphene}, separating coherent $P^{\eta}_{\bf Q} =\langle Y^{\eta \dagger}_{\bf Q}\rangle$ and incoherent hybrid populations $\delta N^{\eta }_{\bf Q} = \langle Y^{\eta\dagger }_{{\bf Q}}Y^{\eta}_{{\bf Q}}\rangle-\langle Y^{\eta \dagger}_{\bf Q}\rangle \langle Y^{\eta}_{\bf Q} \rangle = N^{\eta }_{\bf Q} - |P^{\eta}_{\bf Q}|^2$, leads to the coupled semiconductor Bloch equations
\begin{align}\label{eq:10}
\begin{split}
    i\hbar \partial_t P^\eta_0 &= -(\Ea^\eta_0 + i \Gamma^\eta_0)P^\eta_0 -  \tilde{\mathcal{M}}^\eta_0 \cdot {\bf A}(t)\\[6pt]
    \delta \dot{N}^\eta_{\bf Q} &= \sum_{\xi}{ W^{\xi\eta}_{{\bf 0 Q}}  \abs{P^{\eta}_0}^2 } + \sum_{\xi, {\bf Q'}}{ \left( W^{\xi\eta}_{{\bf Q' Q}} \delta N^\xi_{\bf Q'} - W^{\eta\xi}_{{\bf Q Q'}} \delta N^\eta_{\bf Q} \right) }
\end{split}
\end{align}
 with $W^{\eta\xi}_{{\bf Q Q'}} = \frac{2\pi}{\hbar} \sum_{j,\pm}|\Da^{\eta\xi}_{j,{{\bf Q'-Q}}} |^2 \left( \frac{1}{2} \pm \frac{1}{2} + n^{ph}_{j,{\bf Q'-Q}} \right) \delta \left( \Ea^{\xi}_{\bf Q'} - \Ea^\eta_{\bf Q} \mp   \hbar\Omega_{j{\bf Q'-Q}} \right)$ as the phonon mediated scattering tensor.
 
The large twist angle in the experiment gives rise to very short  moire periods with a length scale comparable with the exciton Bohr radius. Therefore, a strong modification of the exciton center-of-mass motion, i.e. a moire-trapping of excitons is not expected \cite{brem2020tunable} therefore, we neglect the twist angle dependence.

Resonant excitation of the K$_{\rm Mo}$-K$_{\rm Mo}$ exciton leads also to a non-resonant excitation of the K$_{\rm W}$-K$_{\rm W}$ state. The ratio in the exciton occupation of $N_{\rm Mo}/N_{\rm W} \simeq 5$ can be extracted from the experiment. To model this effect in our simulations, we include one main pulse exciting the K$_{\rm Mo}$-K$_{\rm Mo}$ state, and a secondary less intense pulse exciting the K$_{\rm W}$-K$_{\rm W}$ state, imposing the same ratio of the coherent population as in the experiment.

\vspace{.2cm}

\clearpage

\bibliography{bibtexfile}

\end{document}